\documentstyle[eqsecnum,aps]{revtex}
\begin{document}
\draft
\title{$\rho$ - nucleus bound states in Walecka model}
\author{Sanjay K. Ghosh and Byron K. Jennings} 
\address{TRIUMF, 4004 Wesbrook Mall, Vancouver, British Cloumbia, Canada 
V6T 2A3}
\date{\today}
\maketitle
\begin{abstract}
Possible formation of $\rho$ nucleus bound state is studied in the framework of
Walecka model. The bound states are found in different nuclei ranging from 
$^3He$ to $^{208}Pb$. These bound states may have a direct bearing on the 
recent experiments on the photoproduction of $\rho$ meson in the nuclear 
medium.
\end{abstract}

The properties of hadrons in the nuclear medium is a field of current interest. 
These properties at high density and/or temperature are important for the study
of neutron stars and supernovae. The possible explanations of the experimental 
results of heavy ion collisions will also depend on the in-medium behaviour of 
the hadrons. The recent observation of enhanced dilepton production in the 
low invariant mass domain in heavy ion collider experiments \cite{1} has 
triggered speculation \cite{2} that the effective $\rho$- meson mass decreases
in the nuclear medium. The studies using Chiral perturbation theory shows
that even at finite densities there may be a partial restoration of chiral
symmetry leading to the decrease of vector meson masses from their free 
values \cite{3}. 

A number of experiments have been done to study the possible shift of the 
masses of vector mesons $\omega$, $\eta$ \cite{4} and $\rho$ \cite{5}. The most
notable experiments regarding $\rho$ meson mass modification are 
$K^{+} - {^{12}C}$ elastic cross section measurements at 800 MeV that have 
also revealed an enhancement which can be attributed to a shift in $\rho^0$ 
mass \cite{6}; the need for shifted meson masses to explain the spin transfer in
the $\vec {p}$-$A$ scattering experiment at IUCF \cite{7}; and photoproduction 
experiments at the INS facility in Tokyo via the reaction 
$^3He(\gamma, \pi^+, \pi^-)$ at photon energies below the free production 
threshold of 1083 MeV for the $\rho^0$ meson \cite{8,9}. The anlysis of the
experimental data yielded $m^{*}_{\rho^0}$ of 642$\pm$40 MeV/c$^2$ 
and 669$\pm$32 MeV/c$^2$ in the photon energies 800-880 MeV and 880-960 MeV 
respectively\cite{10}. The reanalysis of the single and double $\Delta$ 
experiment \cite{9} gave the $\rho^0$ mass to be 490$\pm$40 MeV/c$^2$ in the
photon energy range 380-700 MeV \cite{11}. The similar trend has been found by
Bianchi {\it et al.} \cite{12} who have developed a model based on hadronic 
fluctuations of the real photon to describe the total photonucleon and 
photonuclear cross sections. A decrease in $\rho$ meson mass in different 
nuclei is found to be necessary to explain the experimental data.

	The substantial change in the $\rho$ meson mass led some authors 
to argue that such large decrease in mass can not be explained by the mean
field picture of the nuclear matter \cite{13}. These authors suggest that 
this decrease in mass should be taken as a signature of the partial 
restoration of chiral symmetry \cite{14}  in ground state nuclei. On the other
hand, Bhattacharyya {\it et al.} \cite{15} showed that the above conclusion is
premature since a proper inclusion of the relevant interactions in a mean 
field description does give a large decrease in the $\rho$ meson mass as 
suggested by the experimental results.  

	In a recent work Popendreou {\it et al.} \cite{16} have shown that the 
reduction in $\rho^0$ meson mass can also be explained through a $\rho^0$-
nucleus bound state. They found that the depth of the potential required to
produce the bound state is consistent with that expected from Dirac 
phenomenology using Brown-Rho scaling. In the present report we have studied 
the formation of $\rho^0$-nucleus bound state using Walecka model. 

	The Walecka model is one of the most popular mean field model for 
nuclear matter \cite{17}. Though there have been lot of modifications \cite{18}
this model still serves as the most effective mean field theory of nuclear 
matter. Starting from the Walecka model Lagrangian \cite{15} one can write 
down the coupled differential equation for the fields in the mean field
approximation. These equations are solved self consistently for different nuclei
to get the density distributions. The $\rho^0$ meson mass is then evaluated 
as the pole of the propagator to get the mass variation with density as well
as radius \cite{15,19}. The evaluation of $\rho^0$ mass involves two
coupling constants. One is the vector coupling of $\rho^0$ meson with
nucleon ($g_{\rho}$) and the other is the tensor coupling of $\rho^0$ with
nucleons ($f_{\rho}$ or $c_{\rho} = f_{\rho}/g_{\rho}$). In general, one can not
fix the value of the tensor coupling constant within the premise of Walecka
model itself. So in practice, one can take $g_{\rho}$ as well as $c_{\rho}$ 
both from some other source like Bonn potential \cite{20} or QCD sum rules 
\cite{21} or keep the $g_{\rho}$ as obtained within Walecka model and use 
$c_{\rho}$ from other calculations \cite{15}. As shown in \cite{15}, the 
reduction in the $\rho^0$ meson mass is different for different parameter sets.
The maximum reduction in mass is obtained from Walecka model parametr set where
as the QCD sum rule parameter set yields the minimum reduction. In the 
following we have used the parameter set $g_{\rho} = 8.912$ as obtained in 
Walecka model by fitting the asymmetric energy for nuclear matter and 
$f_{\rho}=2.866$ \cite{21a} to describe the $\rho$-nucleus bound states. 
Similar values for $f_{\rho}$ are obtained from Bonn potential \cite{20} as 
well as QCD sum rule calculations \cite{21}. The 
average $\rho^0$ masses in $^3He$ for this parameter set are $657$ MeV and 
$600$ MeV without and with tensor interaction respectively.

	The $\rho^0$ meson consists of same flavour quark - antiquarks and as
a result it is not expected to feel the Lorentz Vector potential generated 
by the nuclear environment. The total potential felt by $\rho^0$ is then given 
by $m^{*}_{\rho^0}(r) - m_{\rho^0}$ where $m^{*}_{\rho^0}$ now depends on the 
position from the center of the nucleus. So in a nucleus the static $\rho^0$
meson field $\phi_{\rho}$ is given by,

\begin{equation}
[\nabla^2 + E_{\rho}^2 - m^{*2}_{\rho^0}]\phi_{\rho} = 0
\label{one}.
\end{equation}

To incorporate the width of $\rho^0$ meson in our estimate for the bound states
we assume the phenomenological form as suggested by Saito {\it et al.}
\cite{19}.

\begin{equation}
\tilde{m^{*}_{\rho^0}} = m^{*2}_{\rho^0}(r) - \frac{i}{2} \{ [ m_{\rho^0} -
m^{*}_{\rho^0}(r)]\gamma_{\rho} + \Gamma_{\rho} \} 
\equiv m^{*}_{\rho^0}(r) - \frac{i}{2} \Gamma^{*}_{\rho}(r)
\label{two}
\end{equation}

where $\Gamma_{\rho}$ =150 MeV is the width of $\rho$ meson in free space. 
$\gamma_{\rho}$ is treated as a phenomenological parameter chosen to describe 
the in-medium meson width $\Gamma^{*}_{\rho}$. So we actually solve the 
equation 

\begin{equation}
[\nabla^2 + E_{\rho}^2 - \tilde{m^{*2}}_{\rho^0}(r)]\phi_{\rho}(r) = 0
\label{three}
\end{equation}

The above equation has been solved in the coordinate space using relaxation 
method \cite{22}. This is done in the following way. First we separate the 
eqn.(\ref{three}) in radial and angular parts. Then the wave function 
$\phi_{\rho}$
and energy $E_{\rho}$ are written as $\phi_{\rho} = \phi_{\rho}^{1} + 
i\phi_{\rho}^{2}$ and $E_{\rho} = E_{\rho}^{1} + iE_{\rho}^{2}$, where the 
superscripts $1$ and $2$ denote the real and imaginary parts of the relevant 
quantities. Substituting these in the radial part of the wave equation one gets
two second order coupled differential equations for real and imaginary part of 
the wave function. These are then solved for the real and imaginary part of 
the energy $E_{\rho}$. The single particle energy $E$ can be defined in terms 
of complex eigenenergies $E_{\rho}$ as
\begin{equation}
E_{\rho} = E + m_{\rho} - i \frac{\Gamma}{2}
\end{equation}
The single particle energies $E = Re(E_{\rho} - m_{\rho})$ and the full width
$\Gamma$ for different nuclei are given in Table 1 for two different parameter
sets PI and PII. For PI and PII the vector coupling $g_{\rho}$ is 8.912 whereas 
the tensor coupling $f_{\rho}$ = 2.866 \cite{21a} for PI and $f_{\rho}=0$ for
PII. The table shows that 
a non-zero $\gamma_{\rho}$ increases the width where as the real part remains
almost same. This is evident from eqn. (\ref{two}) which shows that a non-zero
$\gamma_{\rho}$ increases the imaginary part of the potential. Similar effect
has been found in case of $\omega$-nucleus bound states \cite{19} as well. 
The energies in table 1 show that we get very strongly bound $\rho$-nucleus 
states in our present formalism. Without the tensor coupling, the $E = -132.95$
MeV in the $^3He$ in $l=0$ state. The similar value for the $\rho$ bound state
has been found in ref.\cite{16}. On the other hand, we find that $E = -199.14$ 
MeV in $^3He$ in the 
presence of tensor coupling. This high binding is clearly the result of a large
drop in the $\rho^0$ mass in the presence of tensor coupling. In other words,
to get less binding one has to reduce the effect of tensor coupling or 
introduce new effects like form factor which will restrict the dramatic change
in the $\rho^0$ mass.

\begin{table}
\caption{Single particle energy $E$ and full width $\Gamma$ for different 
nuclei for different values of $\gamma_{\rho}$ for two different parameter sets
;  PI $\equiv$ $g_{\rho}=8.912$ and $f_{\rho}=2.866$; PII $\equiv$ $g_{\rho}=
8.912$ and $f_{\rho}=0$}
\begin{tabular}{|ccccccc|}
Nuclei&$\gamma_{\rho}$&l&$E$(PI)&$\Gamma$(PI) &$E$(PII)&$\Gamma$(PII) \\\\
& & &(MeV)&(MeV)&(MeV)&(MeV) \\
\tableline
$^3He$&0&0&-199.14&133.87&-132.95&137.50 \\
 & &1&-126.66&128.48&-68.95&134.02 \\
 &0.2&0&-200.15&177.70&-133.55&168.91 \\
 &   &1&-127.79&163.11&-69.62&156.91 \\
 &0.4&0&-201.32&221.84&-134.09&200.48 \\
 &   &1&-128.93&198.04&-70.05&179.99 \\
\tableline
$^{12}C$&0&0&-232.60&140.5&-162.15&142.82 \\
       & &1&-191.12&133.90&-125.96&137.68 \\
       &0.2&0&-233.18&188.11&-162.45&176.79 \\
       &   &1&-192.00&176.81&-126.43&168.11  \\
       &0.4&0&-233.86&235.89&-162.74&210.83 \\
       &   &1&-193.01&219.96&-126.88&198.64 \\ 
\tableline
$^{40}Ca$&0&0&-254.05&144.92&-181.01&146.14 \\
         & &1&-232.72&140.54&-162.30&142.74 \\
         &0.2&0&-254.36&195.00&-181.16&181.89 \\
         &   &1&-233.26&188.24&-162.58&176.84 \\
         &0.4&0&-254.73&245.17&-181.33&217.68 \\
         &   &1&-233.92&236.11&-162.87&211.01 \\
\tableline
$^{90}Zr$&0&0&-262.44&146.79&-188.35&147.56 \\
         & &1&-249.47&143.77&-176.99&145.23 \\
         &0.2&0&-262.63&197.83&-188.44&183.97\\
         &   &1&-249.84&193.41&-177.17&180.72 \\
         &0.4&0&-262.87&248.93&-188.55&220.40 \\
         &   &1&-250.29&243.17&-177.37&216.25 \\
\tableline
$^{208}Pb$&0&0&-266.14&147.95&-191.59&148.44 \\
          & &1&-258.05&146.07&-184.50&146.99 \\
          &0.2&0&-266.26&199.31&-191.65&185.01 \\
          &   &1&-258.28&196.53&-184.62&182.97 \\
          &0.4&0&-266.41&250.71&-191.72&221.60 \\
          &   &1&-258.57&247.07&-184.76&218.98 \\
\tableline
\end{tabular}
\end{table}

\end{document}